\documentclass[a4paper,12pt]{article}

\newcommand{\sect}[1]{\setcounter{equation}{0}\section{#1}}

\newcommand{\bea}{\begin{eqnarray}}
\newcommand{\eea}{\end{eqnarray}}
\newcommand{\be}{\begin{equation}}
\newcommand{\ee}{\end{equation}}
\newcommand{\vs}[1]{\vspace{#1 mm}}

\renewcommand{\a}{\alpha}

\newcommand{\e}{\epsilon}
\newcommand{\dsl}{\pa \kern-0.5em /}

\newcommand{\pa}{\partial}

\begin{document}
\topmargin 0pt
\oddsidemargin 0mm

\begin{titlepage}
\begin{flushright}
hep-th/0111027\\
SINP-TNP/01-26, OU-HET 399
\end{flushright}

\vs{2}
\begin{center}
{\Large \bf On decoupled theories in (5+1) dimensions 
from\\ (F, D1, NS5, D5) supergravity configuration}
\vs{10}

{\large R.-G. Cai,$^a$\,
I. Mitra,$^b$\, 
N. Ohta $^c$\,
and S. Roy $^b$\,}
\vspace{5mm}

{\em $^a$ Institute of Theoretical Physics\\
Chinese Academy of Sciences, P.O.Box 2735, Beijing 100080, China\\
E-Mail: cairg@itp.ac.cn\\
$^b$ Saha Institute of Nuclear Physics,
 1/AF Bidhannagar, Calcutta-700 064, India\\
E-Mails: indranil, roy@theory.saha.ernet.in\\
$^c$ Department of Physics, Osaka University,
 Toyonaka, Osaka 560-0043, Japan\\
E-Mail: ohta@phys.sci.osaka-u.ac.jp\\}
\end{center}

\vs{5}
\centerline{{\bf{Abstract}}}
\vs{5}
\begin{small}
It is well-known that ($N$, $M$) 5-branes of type IIB supergravity form a
non-threshold bound state with ($N'$, $M'$) strings called the (F, D1, NS5,
D5) bound state where the strings lie along one of the spatial directions of 
the 5-branes (hep-th/9905056). By taking low energy limits in appropriate ways
on this supergravity configuration, we obtain the supergravity duals of various
decoupled theories in (5+1) dimensions corresponding to noncommutative open
string (NCOS) theory, open D-string (OD1) theory and open D5-brane (OD5)
theory. We then study the $SL(2, Z)$ 
transformation properties of these theories. We show that when the asymptotic 
value of the axion ($\chi_0$) is
rational (for which $\chi_0$ can be put to zero), NCOS theory is always 
related to OD1 theory by
strong-weak or S-duality symmetry. We also discuss the 
self-duality conjecture (hep-th/0006062) of both NCOS and OD1 theories. 
On the other hand, when $\chi_0$ is irrational, we find 
that $SL(2, Z)$ duality on NCOS theory gives another NCOS theory 
with different values of the parameters, but for
OD1 theory $SL(2,Z)$ duality always gives an NCOS theory.  $SL(2, Z)$ 
transformation on OD5 theory reveals that it gives rise to 
Little String Theory (LST) when 
$\chi_0$ = rational, but it gives another OD5 theory with different 
values of the parameters when $\chi_0$ is irrational. 
\end{small} 
\end{titlepage}

\newpage
\renewcommand{\thefootnote}{\arabic{footnote}}
\setcounter{footnote}{0}
\setcounter{page}{2}

%%========================section 1 ==========================
\sect{Introduction}
Dynamical theories without gravity appear as particular low energy limits, 
known as decoupling limits, in string/M theory \cite{agmoo,dgk}. These are 
either local
field theories or non-local theories with or without noncommutative 
space-space and/or
space-time structures in diverse dimensions having 16 or
less supercharges and depend upon the kind of string/M theory vacua chosen
and how the low energy limits are taken. Examples are the classic AdS/CFT
correspondence of Maldacena \cite{mal} and their variations \cite{agmoo}. 
Low energy string/M theory
in some sense plays complementary roles to these dynamical theories on the
branes and vice-versa and has been proved to be quite fruitful in the recent
past. These generalized correspondences are believed to shed light on QCD-like
theories and in turn will help us to have the eventual formulation of M theory.

One such correspondence is obtained by considering a particular type IIB 
string theory vacuum consisting of D5-branes in the presence of an electric
field along the brane \cite{gmss}. The corresponding supergravity 
solution is described
by (F, D5) bound state \cite{lurone} where the fundamental strings 
lie along one of the
spatial directions of D5-branes. The low energy limit or the decoupling limit
is taken in such a way that the electric field attains a critical value and
almost balances the tension of the F-strings. The resulting world-volume 
theory on the brane decouples from the bulk closed string modes and becomes a
non-gravitational non-local theory known as the NCOS theory in (5+1) 
dimensions. The (F, D5) supergravity solution in this limit \cite{harone} 
becomes the 
supergravity dual description of NCOS theory\footnote{NCOS theory in (3+1)
dimensions has been obtained in \cite{sst,gmms}.}.

By S-duality of type IIB supergravity (F, D5) solution goes over to
(D1, NS5) solution \cite{mrone,cgnn}, so it is inferred that NS5-branes 
in the presence of a 
near critical RR 2-form field strength would similarly give rise to another
decoupled non-local theory called the OD1 theory in (5+1) dimensions 
\cite{gmss}. Here
also, in the low energy limit, the 2-form field strength attains a critical 
value which almost balances the tension of D-strings and the resulting theory
on the NS5-brane decouples from the bulk closed string modes and contains
fluctuating light open D-strings in its spectra. The (D1, NS5) supergravity
solution in the decoupling limit \cite{mrone,hartwo,alior,ali} describes 
the supergravity dual of OD1 theory.

It is known that these two theories are related by strong-weak duality
symmetry \cite{gmss}, where the coupling constant and 
the length scales in these two
theories are related by $G_o^2 = 1/G_{o(1)}^2$ and $\a'_{\rm eff} = 
G_{o(1)}^2 \tilde{\a}'_{\rm eff}$. By taking further limits $G_o^2 \to
\infty$, $\a'_{\rm eff} \to 0$ with $\a'_{\rm eff} G_o^2\, =$ fixed in
NCOS theory, which amounts to taking $G_{o(1)}^2 \to 0$, 
$\tilde{\a}'_{\rm eff}\,=$ fixed in OD1 theory both these theories reduce to
Little String Theories \cite{brs,sei,dvvone,dvvtwo,lms}. Thus both these 
theories have (5+1) dimensional
Lorentz invariance and reduce to Yang-Mills theory at low energies with
coupling constant $g_{\rm YM}^2 = (2\pi)^3 G_o^2 \a'_{\rm eff} = (2\pi)^3
\tilde{\a}'_{\rm eff}$. In view of this observation, it was conjectured 
\cite{gmss}
that OD1 theory with coupling constant $G_{o(1)}$ and length scale 
$\tilde{\a}'_{\rm eff}$ may be identified with (5+1) dimensional NCOS with
coupling constant $G_o = G_{o(1)}$ and $\a'_{\rm eff} = \tilde{\a}'_{\rm eff}/
G_{o(1)}^2$. Since OD1 and (5+1) NCOS are S-dual to each other, so the above
observation leads to the self-duality conjecture of both the theories.

By applying T-duality on (D1, NS5) solution along all directions transverse
to D-strings and parallel to NS5-branes, it is easy to see that OD1
decoupling limit leads to OD5 decoupling limit \cite{gmss}. In this case, 
an RR 6-form
field strength attains the critical value and almost balances the tension
of D5-branes on NS5-branes leading to OD5 theory. The (NS5, D5) bound state
solution in the decoupling limit describes the supergravity dual 
\cite{mrone,alior,alioz} of OD5 
theory.

In this paper we will consider an $SL(2,Z)$ invariant type IIB string theory
vacuum of the form (F, D1, NS5, D5) which is a non-threshold bound states of
$(N, M)$ 5-branes with the $(N', M')$ strings of type IIB supergravity 
\cite{lurtwo}. The reason for choosing this $SL(2,Z)$ invariant bound
state is that it will lead to the supergravity dual of various decoupled
theories in (5+1) dimensions where the full $SL(2,Z)$ transformation group
of type IIB supergravity can be applied. This is in contrary to the special
case bound states (F, D5) or (D1, NS5) where only the strong-weak duality
behavior can be understood. It should be emphasized that NCOS, OD1 and
OD5 theories obtained from this general supergravity configuration are 
different from those obtained from (F, D5), (D1, NS5) and (NS5, D5)
separately in their field contents and therefore the decoupling limits are
also modified. Also, the self-duality conjecture can be better addressed
in the general $SL(2,Z)$ invariant supergravity solution. The purpose of this
paper is to illustrate the relationships between various decoupled theories
under the general $SL(2,Z)$ transformation of type IIB supergravity from the
gravity point of view.

We
will show here that starting from (F, D1, NS5, D5) bound state 
how all the three decoupling limits discussed above can be
obtained from this single supergravity configuration and study their
$SL(2,Z)$ transformation properties\footnote{$SL(2,Z)$ transformation
on the decoupling limit of (F, D1, D3) bound state has been studied in
\cite{lursone,lurstwo,rsj,co}.}. Note that in this bound state strings
lie along one of the spatial directions of the 5-branes and this solution
preserves half of the space-time supersymmetries of the string theory. The
supergravity solution corresponding to this bound state was constructed in
\cite{lurtwo}. It was mentioned there that the charges of the strings 
and 5-branes
are not independent, but are given as $(N', M') = k(p, q)$ and $(N, M) =
k'(-q, p)$ where $(k, k')$ and $(p, q)$ are two sets of relatively prime
integers. So the number of NS5-branes are $|N| = k'q$ (henceforth we will
call this simply as $N$) and the number of D5-branes are $M = k'p$ in this
bound state. In the special case, when $q = 0$ and $p = 1$, we get (F,
D5) bound state and when $p = 0$, $q = 1$, we get (D1, NS5) 
bound state\footnote{It is also possible to obtain (NS5, D5) bound state
of charges $(-q, p)$ for $k = 0$ and $k' = 1$ and (F, D1) bound state with
four isometries (along $x^2,\, x^3,\, x^4,\, x^5$ directions) and charges
$(p, q)$ for $k' = 0$ and $k = 1$ from this general bound state. However,
because of the charge relation given before, it is not possible to obtain
(F, NS5) and (D1, D5) bound state from here \cite{lurtwo}.}. We
will first briefly review how the (5+1) dimensional NCOS theory and the
OD1-theory are obtained as the decoupling limit on these two special 
bound states. Then we will show that under S-duality the full supergravity
configuration of NCOS theory gets mapped to that of the OD1-theory, if we
identify the parameters of these two theories in a particular way, showing
that these two theories are related by strong-weak duality symmetry. Then we
will take the full (F, D1, NS5, D5) bound state for the case where the
asymptotic value of the axion ($\chi_0$) present in this theory is rational,
which can be put to zero and show how both NCOS-limit and OD1-limit 
can be taken
from this configuration. Even though the field content of the supergravity
configurations are different, we again find that under similar
identification of parameters, S-dual NCOS theory gets mapped to OD1-theory.
So, the same conclusion holds even for this more general NCOS theory and OD1
theory. One of the motivations for considering NCOS and OD1 limit for this
more general solution was to see whether we can understand the self-duality
conjecture\footnote{Indeed we have seen in \cite{mrtwo} (see also
\cite{lu,lars}) that the self-duality of
OD3-theory can be shown from an $SL(2,Z)$ invariant bound state
configuration (NS5, D5, D3) of type IIB supergravity. We obtained an OD3
decoupling limit for this solution and have shown that under S-duality, it
gives rise to another OD3 theory with the same length scale as the original
OD3-theory and the coupling constants are related as $\hat{G}_{o(3)} =
1/G_{o(3)}$.} of both (5+1) dimensional NCOS and OD1-theory mentioned
earlier. As we have already mentioned, the self duality of either the (5+1)
dimensional NCOS theory or OD1 theory means that we should be able to
identify the OD1 theory with coupling constant $G_{o(1)}$ and length scale 
$\tilde{\a}'_{\rm eff}$ with the NCOS theory with coupling constant
 $G_o = G_{o(1)}$ and length scale 
$\a'_{\rm eff} = \tilde{\a}'_{\rm eff}/G^2_{o(1)}$. If we can
identify the corresponding supergravity duals of these two theories with the
above identification, then it will lend support to the self-duality
conjecture \cite{gmss}. We will show that under appropriate 
identification of the
parameters along with a condition the two supergravity configuration can
indeed be identified. However, a closer look at some of the conditions
required for the identification reveals that it can only be made in the
range of the energy parameter in the NCOS theory where the dilaton blows up,
i.e. where supergravity description breaks down. This may be an indication
that the
self-duality conjecture of either of these theories cannot be tested  at the
level of supergravity dual and some non perturbative techniques may be
required.

We then study the general $SL(2,Z)$ transformation property of both NCOS
theory and the OD1 theory for the $SL(2,Z)$ invariant configuration when the
asymptotic value of the axion is irrational. Here we
find that under a generic $SL(2,Z)$ transformation an NCOS theory always
gives another NCOS theory with different values of the parameters. We give
relations of the coupling constant and the length scales of the $SL(2,Z)$
transformed NCOS theory in terms of those of the original NCOS theory.
However, the story is different for OD1-theory. For OD1 theory we find that
even for the generic $SL(2,Z)$ transformation OD1-theory reduces to NCOS
theory with rational $\chi_0$. Since NCOS theory with rational $\chi_0$ is
mapped to OD1 theory with rational $\chi_0$ under S-duality, so it shows that
OD1 theory with irrational $\chi_0$ is equivalent to the same theory with
rational $\chi_0$. So $SL(2,Z)$ transformation in this case does not give
any new information.

Finally we also obtain an OD5-limit in the general $SL(2,Z)$ invariant
supergravity
configuration (F, D1, NS5, D5). Here we find that for rational $\chi_0$, OD5
theory goes over to Little String Theory under S-duality of type IIB
supergravity. However for irrational $\chi_0$, the general $SL(2,Z)$
transformation gives another OD5 theory with different values of the
parameters. The explicit relations between the parameters in the two OD5
theories are given.

The organization of the paper is as follows. In section 2, we give the
supergravity configuration of (F, D1, NS5, D5) bound state. We briefly
review the NCOS limit and the OD1-limit for the special cases of (F, D5) and
(D1, NS5) configuration in section 3. Here we also discuss the S-duality
between NCOS and OD1-theories. In section 4, both NCOS and OD1-limits are
obtained for the general (F, D1, NS5, D5) bound state when $\chi_0$ is
rational. We discuss the S-duality as well as the self-duality of either NCOS
or OD1 theory at the level of supergravity dual. The general
$SL(2,Z)$ transformations when $\chi_0$ is irrational is discussed in
section 5. The OD5-limit and its $SL(2,Z)$ transformation for both the cases
when $\chi_0$ is rational as well as irrational is discussed 
in section 6. Our conclusion is presented in section 7.                 

\sect{The (F, D1, NS5, D5) bound state}  

The supergravity configuration of this  bound state was
constructed in \cite{lurtwo}. Note that the metric in \cite{lurtwo} was 
written in
Einstein-frame which asymptotically becomes Minkowskian. Also the solution
was written for $\chi_0 = 0$. We here rewrite the solution with the metric in
string frame such that the string frame metric becomes asymptotically
Minkowskian with an appropriate scaling of the coordinates. Also, we write
the solution for $\chi_0\neq0$, by an $SL(2,R)$ transformation on the
solution of \cite{lurtwo}. The configuration is, 
\begin{eqnarray}
&& ds^2 =H'^{1/2}H''^{1/2}\left[H^{-1}(-(d x^0)^2 + (dx^1)^2) +H'^{-1}
\sum_{i=2}^5(dx^i)^2
    + dr^2 + r^2 d\Omega_3^2\right], \nonumber \\
&& e^{\phi}=g_s H^{-1/2}H'' ,\ \ \  \chi =\frac{
    \sin\varphi \cos\varphi(H-1)}{g_s H''} + \chi_0 =
\frac{\tan\varphi(1-H''^{-1})}{g_s}+\chi_0, \nonumber\\
&& F_{NS}= \sin\varphi\sin\psi dH^{-1} \wedge d x^0 \wedge dx^1
   + 2 \cos\psi\cos\varphi\, Q_5 \,\epsilon_3,
   \nonumber \\
&& F_{RR}= - \left(\chi_0 - \frac{\cot\varphi}{g_s}\right)
     \sin\psi\sin\varphi\,dH^{-1}\wedge dx^0 \wedge dx^1 -
2 \cos\psi\cos\varphi \left(\chi_0 + \frac{\tan\varphi}{g_s}\right)
Q_5\,\epsilon_3,
\nonumber \\
&& C_4 =  \frac{\tan\psi}{g_s} \left(1 -
   H'^{-1}\right) d x^2\wedge d x^3 \wedge d x^4 \wedge d x^5.
\label{eq:twone}
\end{eqnarray}
Note from (\ref{eq:twone}) that the strings lie along $x^1$ direction and the
5-branes lie along $x^1, x^2, x^3, x^4, x^5$ directions. Also $r
=\sqrt{x_{6}^2+x_{7}^2+x_{8}^2+x_{9}^2}$, $d\Omega_{3}^2 =
d\theta^2+\sin^2\theta d\phi_{1}^2+\sin^2\theta\sin^2\phi_{1} d\phi_{2}^2
$ is the line element of the unit 3-sphere transverse to the 5-branes and
 $\e_3$ is its volume form. $g_{s} = e^{\phi_0}$ is the string coupling
constant, $\chi$ is the RR scalar (axion), $F_{NS}$ and $F_{RR}$ are
respectively the NSNS and RR 3-form field strength. $C_4$ is the RR
4-form gauge field whose field strength is self dual. The Harmonic functions
$H$, $H'$, $H''$ are given as,
\begin{eqnarray}
H &=& 1 + \frac{Q_5}{r^2},\nonumber\\
H' &=& 1 + \frac{\cos^2\psi Q_5}{r^2},\nonumber\\
H'' &=& 1 + \frac{\cos^2\varphi Q_5}{r^2},
\label{eq:twtwo}
\end{eqnarray}
where the angles are defined as,
\begin{equation}
\cos\varphi = \frac{q}{\sqrt{(p-\chi_0 q)^2 g_s^2 + q^2}}, \qquad
\cos\psi = \frac{k'}{\sqrt{k^2 g_{s}^2
 + k'^2 }}.
\label{eq:twthree}
\end{equation}
Here $(p, q)$ and $(k, k')$ are two sets of relatively prime integers 
which appear
in the integral charges of (F, D1, NS5, D5) system as $(N',M',N,M) =
(kp,kq,-k'q,k'p)$. Note that the number of NS5 branes is $|N| = k'q$ and D5
branes is $M = k'p$. The form of $Q_5$ in (\ref{eq:twtwo}) is given as 
\begin{equation}
Q_5 = \left[(p-\chi_0 q)^2 g_s^2 + q^2\right]^{1/2}
\left[k^2 g_{s}^{2} + k'^2 \right]^{1/2} \alpha'.
\label{eq:twfour}
\end{equation}
Note that $Q_5$ can be expressed in terms of the angles as follows
\begin{equation}
Q_5 = \frac{N\alpha'}{\cos\varphi\cos\psi} = \frac{\left(M-\chi_{0}N\right)
g_{s}\alpha'}{\sin\varphi\cos\psi}. 
\label{eq:twfive}
\end{equation}
We will also use these forms of $Q_5$ while discussing OD1 and NCOS limits
later. We note here that among the angles given in (\ref{eq:twthree}), 
$\cos\psi$ is
$SL(2,Z)$ invariant, but $\cos\varphi$ is not. Also $Q_5$ is $SL(2,Z)$
invariant. Therefore while the harmonic functions $H$ and $H'$ are $SL(2,Z)$
invariant $H''$ is not. The RR 4-form in (\ref{eq:twone}) is $SL(2,Z)$
invariant but the other fields change under $SL(2,Z)$ transformation. These
will be useful when we study the $SL(2,Z)$ transformation of various
decoupled theories.

\sect{NCOS and OD1 limit in the special cases and S-duality}

We have already mentioned in the introduction that we can obtain (F, D5)
bound state and (D1, NS5) bound state as special cases from the general
bound state (F, D1, NS5, D5) given in the previous section.
It is clear that when the integers $q = 0$, $p = 1$ and $\chi_0 = 0$, 
we get (F, D5) bound
state with charges $(k,k')$ and similarly when $q = 1, p = 0$ along with
$\chi_0 = 0$ we get (D1,
NS5) bound state with charges $(-k',k)$. We will review the NCOS and OD1
decoupling limits for (F, D5) and (D1, NS5) bound states repectively and
show that the supergravity dual of NCOS theory gets mapped to that of OD1
theory by S-duality. This has already been discussed in \cite{hartwo}, 
but we will
show the full mapping including the NSNS and RR gauge fields.

\subsection{(F, D5) solution and NCOS-limit}

When $q = 0$, $p = 1$ and $\chi_0 = 0$ we find from (\ref{eq:twthree}) that 
$\cos\varphi =
0$. Therefore, $H'' = 1$. 
In this case
(F, D1, NS5, D5) configuration in (\ref{eq:twone}) reduces to (F, D5)
solution and is given as, 
\begin{eqnarray}
&& ds^2 =H'^{1/2}\left[H^{-1}(-(d x^0)^2 + (dx^1)^2) +H'^{-1}
\sum_{i=2}^5(dx^i)^2 
+ dr^2 + r^2 d\Omega_3^2\right], \nonumber \\
&& e^{\phi}=g_s H^{-1/2} ,\ \ \  \chi = 0, \nonumber \\
&& F_{NS}= \sin\psi dH^{-1} \wedge d x^0 \wedge dx^1,
   \nonumber \\
&& F_{RR}= - \frac{2\cos\psi}{g_s}
Q_5\,\epsilon_3,
\nonumber \\
&& C_4 =  \frac{\tan\psi}{g_s} \left(1 -
   H'^{-1}\right) d x^2\wedge d x^3 \wedge d x^4 \wedge d x^5.
\label{eq:thone}
\end{eqnarray}
From the second expression of (\ref{eq:twfive}) we find the form of $Q_5$ as 
\begin{equation}
Q_5 = \frac{M g_{s}\alpha'}{\cos\psi} 
\label{eq:thtwo}
\end{equation}
and so the harmonic functions have the forms (see eq.(\ref{eq:twtwo}))  
\begin{eqnarray}
H &=& 1 + \frac{M g_{s}\alpha'}{\cos\psi r^2},\nonumber\\
H' &=& 1 + \frac{\cos\psi M g_{s}\alpha'}{r^2},
\label{eq:ththree}
\end{eqnarray}
where $M$ is the number of D5-branes. The NCOS decoupling limit \cite{gmss} can be
summarized as follows. We define a positive, dimensionless scaling parameter
$\epsilon$ and take $\cos\psi = \epsilon\rightarrow 0$, keeping the
following quantities fixed,
\begin{equation}
\alpha'_{\rm eff} = \frac{\alpha'}{\epsilon}, \qquad  G_{o}^2 = 
\epsilon g_{s}, \qquad  
u = \frac{r}{{\sqrt{\epsilon}}\alpha'_{\rm eff}}. 
\label{eq:thfour}
\end{equation}
In the above $G^{2}_o$ is the coupling constant, $\alpha'_{\rm eff}$ is the
length scale and $u$ is the energy parameter in the NCOS theory. Under this
limit, the harmonic functions in (\ref{eq:ththree}) reduce to  
\begin{equation}
H = \frac{1}{a^2\epsilon^2 u^2}, \qquad
H' = \frac{h'}{a^2 u^2},
\label{eq:thfive}
\end{equation} 
where we have defined $a^2 = \alpha'_{\rm eff}/MG_{o}^2$ and $h' = 1+a^2 u^2$.
Then the supergravity configuration in (\ref{eq:thone}) take the forms 
\begin{eqnarray}
&& ds^2 =\epsilon h'^{1/2}au\left[-(d \tilde x^0)^2 
+ (d\tilde x^1)^2 +h'^{-1}
\sum_{i=2}^5(d\tilde x^i)^2 
+ \frac{M G_{o}^2\alpha'_{\rm eff}}{u^2}
\left(du^2 + u^2 d\Omega_3^2\right)\right], \nonumber \\
&& e^{\phi}= G_{o}^2 au ,\nonumber\\
&& B_{01}= \epsilon a^2 u^2,\nonumber\\
&& F_{RR}= - 2\epsilon M\alpha'_{\rm eff}\epsilon_{3},\nonumber\\
&& C_{2345} =\epsilon^2 /(G_{o}^2 h').
\label{eq:thsix}
\end{eqnarray}
In the above we have defined the fixed coordinates as 
\begin{equation}
\tilde x^{0,1} = \sqrt\epsilon x^{0,1}, \qquad
\tilde x^{2,\ldots,5} = \frac{1}{\sqrt\epsilon} x^{2,\ldots,5}.
\label{eq:thseven}
\end{equation}
Note that this is precisely the (5+1) dimensional NCOS limit given in
\cite{gmss}. Also $B_{01}$ in (\ref{eq:thsix}) is the NSNS 2-form potential in
component form and $C_{2345}$ is the RR 4-form potential in component form.
(\ref{eq:thsix}) describes the supergravity dual of (5+1) dimensional NCOS
theory. The gravity dual description is valid as long as $e^{\phi}\ll 1$
and the curvature measured in units of $\alpha'$ remains small. 

\subsection{(D1, NS5) solution and OD1 limit}

The (D1, NS5) solution can be obtained from (\ref{eq:twone}) 
when $q = 1,\, p =
0$ and also $\chi_0 = 0$. In this case $\cos\varphi = 1$ 
(from (\ref{eq:twthree})) and
therefore from (\ref{eq:twtwo}) harmonic functions $H = H''$. This solution
now takes the form,
\begin{eqnarray}
&& d\bar{s}^2 =H'^{1/2}H^{1/2}\left[H^{-1}(-(d x^0)^2 + (dx^1)^2) +H'^{-1}
\sum_{i=2}^5(dx^i)^2 
+ dr^2 + r^2 d\Omega_3^2\right], \nonumber \\
&& e^{\bar{\phi}}=g_s H^{1/2} ,\ \ \  \bar{\chi} = 0, \nonumber \\
&& \bar{F}_{NS}= 2\cos\psi Q_{5}\epsilon_{3},
   \nonumber \\
&& \bar{F}_{RR}=  \frac{\sin\psi}{g_s} dH^{-1}\wedge d x^0 \wedge d x^1,
\nonumber \\
&& \bar{C}_4 = \frac{\tan\psi}{g_{s}} \left(1 -
   H'^{-1}\right) d x^2\wedge d x^3 \wedge d x^4 \wedge d x^5.
\label{eq:theight}
\end{eqnarray}
To avoid notational confusion we have denoted the fields of (D1, NS5) solution
with a `bar'.
Now from the first expression of (\ref{eq:twfive}) we have the form 
of $Q_{5}$ as
\begin{equation}
Q_5 = \frac{N \alpha'}{\cos\psi}
\label{eq:thnine}
\end{equation}
and therefore the harmonic functions in (\ref{eq:twtwo}) are given as
\begin{equation}
H = 1+\frac{N\alpha'}{\cos\psi r^2},\qquad
H' = 1+\frac{\cos\psi N\alpha'}{r^2}
\label{eq:thten}
\end{equation}
where $N$ is the number of NS5-branes. The OD1 decoupling limit is obtained by
defining the positive scaling parameter $\epsilon$ and taking $\cos\psi =
\epsilon\rightarrow 0$, keeping the following quantities fixed
\begin{equation}
\tilde\alpha'_{\rm eff} = \frac{\tilde\alpha'}{\epsilon},\qquad 
 G_{o(1)}^2 = \frac{g_{s}}{\epsilon}, \qquad  
\tilde{u} = \frac{r}{\epsilon\tilde{\alpha}'_{\rm eff}} 
\label{eq:theleven}
\end{equation}
where $\tilde\alpha' = \alpha'/G^2_{o(1)}$. In the above
$\tilde\alpha'_{\rm eff}$ is the length scale, $G^2_{o(1)}$ is the coupling
constant and $\tilde u $ is the energy parameter in the OD1 theory. In this
limit the harmonic functions in (\ref{eq:thten}) are given as
\begin{equation}
H = \frac{1}{c^2 \epsilon^2 \tilde u^2},\qquad
H' = \frac{f'}{c^2 \tilde u^2}
\label{eq:thtwelve}
\end{equation}
where we have defined $c^2 = \tilde\alpha'_{\rm eff}/(N G^2_{o(1)})$ and $f' =
1+c^2\tilde u^2$. The supergravity configuration in (\ref{eq:theight}) then
takes the form,  
\begin{eqnarray}
&& d\bar s^2 =\epsilon G^2_{o(1)} f'^{1/2}\left[-(d \bar x^0)^2 
+ (d\bar x^1)^2 +f'^{-1}
\sum_{i=2}^5(d\bar x^i)^2 
+ \frac{N \tilde{\alpha}'_{\rm eff}}{\tilde u^2}
\left(d\tilde u^2 + \tilde u^2 d\Omega_3^2\right)\right], \nonumber \\
&& e^{\bar\phi}= \frac{G_{o(1)}^2}{c\tilde u} ,\nonumber\\
&& \bar F_{NS}= 2\epsilon N G^2_{o(1)}\tilde\alpha'_{\rm eff}
\epsilon_3\nonumber\\
&& \bar C_{01}= \epsilon c^2\tilde u^2,\nonumber\\
&& \bar C_{2345} =\epsilon^2 G_{o(1)}^2/ f'.
\label{eq:ththirteen}
\end{eqnarray}
In the above we have defined the fixed coordinates as, 
\begin{equation}
\bar x^{0,1} = \frac{x^{0,1}}{G_{o(1)}},\qquad
\bar x^{2,\ldots,5} = \frac{1}{\epsilon G_{o(1)}} x^{2,\ldots,5}.
\label{eq:thfourteen}
\end{equation}
Here $\bar C_{01}$ is the RR 2-form potential in the component
form and $\bar{C}_{2345}$ is the RR 4-form potential in component form.
Eq.(\ref{eq:ththirteen}) describes the supergravity dual of OD1 theory.

\subsection{S-duality between NCOS and OD1 theory}

In this subsection we will show that under the S-duality of type IIB
supergravity, the gravity dual configuration of NCOS theory gets mapped to
those of OD1 theory. Note that the S-dual configuration of NCOS theory would
be given as
\begin{eqnarray}
&& d\hat s^2 = e^{-\phi}ds^2,\nonumber\\ 
&& e^{\hat\phi}= e^{-\phi},\nonumber\\
&& \hat F_{NS}= - F_{RR},\nonumber\\
&& \hat C_{01}= B_{01},\nonumber\\
&& \hat C_{2345} = C_{2345}.
\label{eq:thfifteen}
\end{eqnarray}
Using these we get from (\ref{eq:thsix}) the S-dual supergravity configuration
of NCOS theory as,
\begin{eqnarray}
&& d\hat s^2 =\frac{\epsilon h'^{1/2}}{G^2_o}\left[-(d \tilde x^0)^2 
+ (d\tilde x^1)^2 +h'^{-1}\sum_{i=2}^5
(d\tilde x^i)^2 
+ \frac{M G_{o}^2\alpha'_{\rm eff}}{u^2}
\left(du^2 + u^2 d\Omega_3^2\right)\right], \nonumber \\
&& e^{\hat\phi}= \frac{1}{G_{o}^2 au} ,\nonumber\\
&& \hat F_{NS}= 2\epsilon M \alpha'_{\rm eff} \epsilon_3,\nonumber\\
&& \hat C_{01}= \epsilon a^2 u^2,\nonumber\\
&& \hat C_{2345} =\epsilon^2 /(G_{o}^2 h').
\label{eq:thsixteen}
\end{eqnarray}
Here $\tilde x^{0,1}$ and $\tilde x^{2,\ldots,5}$ are the same as given in
(\ref{eq:thseven}). 
Comparing (\ref{eq:thsixteen}) with the field in OD1 theory given in
(\ref{eq:ththirteen}) we find that they match exactly with the 
following identification
\bea
&& h' = f', \qquad G^2_{o} = \frac{1}{G^2_{o(1)}},\qquad \alpha'_{\rm eff} =
G^2_{o(1)}\tilde\alpha'_{\rm eff},\qquad M  = N,   \nonumber\\
&& {\rm and} \quad   \tilde x^{0,1}=\bar x^{0,1},\qquad 
\tilde x^{2,\ldots,5}=\bar x^{2,\ldots,5}.
\label{eq:thseventeen}
\eea
The first condition in (\ref{eq:thseventeen}) implies
\begin{equation}
a^2 u^2 = c^2\tilde u^2 \quad \Rightarrow \quad u=\frac{\tilde u}{G^3_{o(1)}}.
\label{eq:theighteen}
\end{equation}
This gives a relation between the energy parameters in NCOS theory and OD1
theory. Thus we find that the coupling constant and the length scales in
these two theories are related as given in (\ref{eq:thseventeen}) showing 
that these
two theories are S-dual to each other. From the coordinate relation
given in (\ref{eq:thseventeen}), it is clear that the coordinates of the
gravity dual configurations of NCOS theory ($x^{0,1,\ldots,5}({\rm NCOS}))$
and OD1 theory ($x^{0,1,\ldots,5}({\rm OD1}))$ have a relative scaling of
the form $x^{0,1,\ldots,5}({\rm NCOS}) = 1/(G_{o(1)}\sqrt{\e}) 
x^{0,1,\ldots,5}({\rm OD1})$. Similarly, the radial coordinates in these
two theories also have a relative scaling $r ({\rm NCOS}) = 1/(G_{o(1)}
\sqrt{\e}) r ({\rm OD1})$ as can be seen from (\ref{eq:thfour}) and
(\ref{eq:theleven}) if the parameters in these two theories are related as in
(\ref{eq:thseventeen}) and (\ref{eq:theighteen}). The reason for this
can be traced back as follows. Note that the metric for both (F, D5) and
(D1, NS5) supergravity configuration given respectively in (\ref{eq:thone})
and (\ref{eq:theight}) are asymptotically Minkowskian and we have taken
decoupling limits on them. However, when we take S-duality on the gravity dual
of NCOS in (\ref{eq:thfifteen}), the S-dual metric ($d\hat{s}^2$) is not
asymptotically Minkowskian, since, as usual, we have absorbed a factor of
$g_s$ in $ds^2$ in the r.h.s. of the metric expression in (\ref{eq:thfifteen}).
So, to compensate this effect we have to multiply a $g_s^{-1/2} = 
\sqrt{\e}/G_o (= \sqrt{\e} G_{o(1)})$ on the NCOS coordinate while the
OD1 coordinates remain the same. 
We note by comparing 
the gravity dual
configurations of NCOS theory in (\ref{eq:thsix}) and OD1 theory in
(\ref{eq:ththirteen}) that they cannot be identified with the parametric 
relations
$G_{o} = G_{o(1)}$ and $\alpha'_{\rm eff} = \tilde\alpha'_{\rm
eff}/G^2_{o(1)}$ and so the self-duality conjecture of (5+1) dimensional
NCOS and OD1 cannot be tested at the level of gravity dual. We will mention
more about it in the context of full (F, D1, NS5, D5) solution in the next
section.

\sect{NCOS and OD1 limit for (F, D1, NS5, D5) bound state with rational
$\chi_0$ and S-duality}

In this section we consider the full (F, D1, NS5, D5) supergravity
configuration with rational $\chi_0$. When $\chi_0$ is rational we can
always make an $SL(2,Z)$ transformation such that $\chi_0$ vanishes.
Therefore, we will take the supergravity configuration (\ref{eq:twone})
with $\chi_0 = 0$. We will obtain both the NCOS limit and OD1 limit for this
configuration discussed in the previous section. We will show as in the
previous section that the supergravity dual of NCOS and OD1 get mapped to
each other under S-duality. We will also comment on the self-duality of
(5+1) dimensional NCOS theory.

\subsection{(F, D1, NS5, D5) solution and NCOS limit}

The NCOS limit for the (F, D5) solution is given in (\ref{eq:thfour}). In that
case $\cos\varphi = 0$, but here we will take $\cos\varphi$ to scale as
\begin{equation}
\cos\varphi = l\epsilon,
\label{eq:foone}
\end{equation}
where $l$ is a finite parameter such that $l\epsilon\rightarrow 0$, but $|l|$
obviously has to be less than $1/\epsilon$. So eq.(\ref{eq:thfour}) along with
(\ref{eq:foone}) define the NCOS limit in this case. The harmonic functions
(\ref{eq:twtwo}) in this case take the forms,   
\begin{eqnarray}
H &=& 1 + \frac{M g_{s}\alpha'}{\cos\psi\sin\varphi r^2},\nonumber\\
H' &=& 1 + \frac{\cos\psi M g_{s}\alpha'}{\sin\varphi r^2},\nonumber\\
H''&=& 1 + \frac{\cos^2\varphi M g_{s}\alpha'}{\cos\psi
\sin\varphi r^2},
\label{eq:fotwo}
\end{eqnarray}
where we have used the form of $Q_5$ in (\ref{eq:twfive}). Here $M = k'p$ 
denotes the number of D5-branes. Now in the decoupling limit (\ref{eq:thfour})
and (\ref{eq:foone}) the harmonic functions reduce to
\begin{equation}
H = \frac{1}{a^2\epsilon^2 u^2},\qquad
H' = \frac{h'}{a^2 u^2},\qquad
H''= \frac{h''}{\tilde a^2 u^2}.
\label{eq:fothree}
\end{equation}
In the above $a^2 = \alpha'_{\rm eff}/(MG^2_o)$ as in (\ref{eq:thfive}) and
$h' = 1 + a^2 u^2$, but $\tilde a^2 = \alpha'_{\rm eff}/(MG^2_o l^2)$ and $h''
= 1 + \tilde a^2 u^2$. We thus have $a /\tilde a = l$ and $a\tilde a
= \alpha'_{\rm eff}/(MG^2_o l)$. Using the decoupling limit and 
eq.(\ref{eq:fothree}), we find that the full supergravity configuration 
(eq.(\ref{eq:twone}) with $\chi_0 = 0$) reduce to,
\begin{eqnarray}
&& ds^2 =\epsilon l h'^{1/2}h''^{1/2}\left[-(d \tilde x^0)^2 
+ (d\tilde x^1)^2 +h'^{-1}
\sum_{i=2}^5(d\tilde x^i)^2 
+ \frac{M G_{o}^2\alpha'_{\rm eff}}{u^2}
\left(du^2 + u^2 d\Omega_3^2\right)\right], \nonumber \\
&& e^{\phi}= G_{o}^2 h''\frac{l^2}{au}=\frac{G^2_o h''l}{\tilde a u}
 ,\qquad \chi=\frac{1}{lh''G^2_o}, \nonumber\\
&& B_{01}= \epsilon a^2 u^2, \qquad F_{NS}=2\epsilon lMG^2_o\alpha'_{\rm eff}
\epsilon_3,\nonumber\\
&& C_{01}=\epsilon^3\frac{l}{G^2_o}a^2 u^2 , \qquad 
F_{RR}= - 2\epsilon M\alpha'_{\rm eff}\epsilon_{3},\nonumber\\
&& C_{2345} =\epsilon^2 /(G_{o}^2 h').
\label{eq:fofour}
\end{eqnarray}
In the above the fixed coordinates $\tilde x^{0,1,\ldots,5}$ are again as given
before in (\ref{eq:thseven}). Note from eq.(\ref{eq:twone}) that NSNS and RR
3-form field strengths have two parts each, so in (\ref{eq:fofour}) $B_{01}$
and $C_{01}$ denote respectively the 01 components of NSNS and RR 2-form
potentials. The other parts $F_{NS}$ and $F_{RR}$ are kept as it is. 
$C_{2345}$ is the 4-form
potential in component form. Eq.(\ref{eq:fofour}) represents the gravity
dual description of NCOS theory.

\subsection{(F, D1, NS5, D5) solution and OD1 limit}

The OD1 limit for the (D1, NS5) solution is given in (\ref{eq:theleven}). In
that case the angle $\cos\varphi = 1$ i.e. $\sin\varphi = 0$, but here we will
take $\sin\varphi$ to scale as, 
\begin{equation}
\sin\varphi = \tilde l\epsilon,
\label{eq:fofive}
\end{equation}
where $\tilde l$ is another finite parameter such that $\tilde
l\epsilon\rightarrow 0$, with $|\tilde l| <1/\epsilon$. So, 
eq.(\ref{eq:theleven})
along with (\ref{eq:fofive}) define the OD1-limit in this case. The harmonic
functions (\ref{eq:twtwo}) for this case take the forms,
\begin{eqnarray}
H &=& 1 + \frac{N\alpha'}{\cos\psi\cos\varphi r^2},\nonumber\\
H' &=& 1 + \frac{\cos\psi N\alpha'}{\cos\varphi r^2},\nonumber\\
H''&=& 1 + \frac{\cos\varphi N\alpha'}{\cos\psi
 r^2},
\label{eq:fosix}
\end{eqnarray}
where we have used the form of $Q_5$ in eq (\ref{eq:twfive}). Here $N = k'q$
represents the number of NS5-branes. In the OD1 decoupling limit given by 
eq.(\ref{eq:theleven}) and (\ref{eq:fofive}), the harmonic functions in
(\ref{eq:fosix}) reduce to 
\begin{equation}
H = \frac{1}{c^2\epsilon^2 \tilde u^2},\qquad
H' = \frac{f'}{c^2 \tilde u^2},\qquad
H''= \frac{1}{c^2\epsilon^2 \tilde u^2}.
\label{eq:foseven}
\end{equation}
In the above $c^2 = \tilde\alpha'_{\rm eff}/(NG^2_{o(1)})$ as 
given before and
$f' = 1 + c^2 \tilde u^2$. Using this, the full supergravity configuration
(eq.(\ref{eq:twone}) with $\chi_0 = 0$) takes the form, 
\begin{eqnarray}
&& d\bar s^2 =\epsilon f'^{1/2}G^2_{o(1)}\left[-(d \bar x^0)^2 
+ (d\bar x^1)^2 +f'^{-1}\sum_{i=2}^5
(d\bar x^i)^2 
+ \frac{N\tilde\alpha'_{\rm eff}}{\tilde u^2}
\left(d\tilde u^2 + \tilde u^2 d\Omega_3^2\right)\right], \nonumber \\
&& e^{\bar\phi}= \frac{G^2_{o(1)}}{c\tilde u}
 ,\qquad \bar\chi=\frac{\tilde l}{G^2_{o(1)}}, \nonumber\\
&&\bar B_{01}=\epsilon^3 \tilde{l} G^2_{o(1)}c^2 \tilde u^2 , \qquad 
\bar F_{NS}= 2\epsilon NG^2_{o(1)}\tilde\alpha'_{\rm eff}
\epsilon_{3},\nonumber\\
&& \bar C_{01}= \epsilon c^2 \tilde u^2, \qquad 
\bar F_{RR}= -2\epsilon \tilde l N\tilde{\alpha}'_{\rm eff}
\epsilon_3,\nonumber\\
&& \bar{C}_{2345} =\epsilon^2 G_{o(1)}^2/f'.
\label{eq:foeight}
\end{eqnarray}
Here the fixed coordinates, $\bar x^{0,1,\ldots,5}$ are the same as defined
before in (\ref{eq:thfourteen}). So  eq.(\ref{eq:foeight}) represents 
the gravity
dual description of OD1 theory.

\subsection{S-duality between NCOS and OD1 theory in general case}

We have obtained the gravity dual configurations of NCOS and OD1 theory from
the decoupling limits on the same supergravity solution (F, D1, NS5, D5) in
eqs (\ref{eq:fofour}) and (\ref{eq:foeight}). In this subsection we will show
that these two theories are related by S-duality even for these more general
NCOS and OD1 theories. The S-dual configuration of NCOS theory would be
given as
\begin{eqnarray}
&& d\hat s^2 = |\lambda|ds^2,\nonumber\\ 
&& e^{\hat\phi}= |\lambda|^2e^{\phi},\qquad \hat \chi= 
-\frac{\chi}{|\lambda|^2},\nonumber\\
&& \hat B_{01}= -C_{01}, \qquad \hat F_{NS}= -F_{RR},\nonumber\\
&& \hat C_{01}= B_{01},\qquad \hat F_{RR}= F_{NS},\nonumber\\
&& \hat C_{2345} =C_{2345},
\label{eq:fonine}
\end{eqnarray}
where $|\lambda| = \sqrt{\chi^2 + e^{-2\phi}}$. 
As before here also we have absorbed a factor of $|\lambda_0|^{-1}$
in $ds^2$ on the r.h.s. of the metric expression in (\ref{eq:fonine}). 
The value of $|\lambda|$ can
be calculated from (\ref{eq:fofour}) and has the form $|\lambda| =
1/(l G_o^2 h''^{1/2})$.

So, using (\ref{eq:fonine}) and (\ref{eq:fofour}), we find the S-dual
configuration of NCOS theory as,
\begin{eqnarray}
&& d\hat s^2 =\frac{\epsilon h'^{1/2}}{G^2_o}\left[-(d \tilde x^0)^2 
+ (d \tilde x^1)^2 +h'^{-1}\sum_{i=2}^5
(d\tilde x^i)^2 
+ \frac{M G_{o}^2\alpha'_{\rm eff}}{u^2}
\left(du^2 + u^2 d\Omega_3^2\right)\right], \nonumber \\
&& e^{\hat\phi}= \frac{1}{G_{o}^2 au}, \qquad \hat\chi= -l G^2_o,\nonumber\\
&& \hat B_{01}= -\epsilon^3\frac{l}{G^2_o}a^2 u^2, \qquad
\hat F_{NS}= 2\epsilon M \alpha'_{\rm eff} \epsilon_3,\nonumber\\
&& \hat C_{01}= \epsilon a^2 u^2, 
\qquad \hat F_{RR}= 2\epsilon lMG^2_o\alpha'_{\rm eff}\epsilon_3,\nonumber\\
&& \hat C_{2345} = \frac{\epsilon^2}{G_{o}^2 h'}.
\label{eq:foten}
\end{eqnarray}
By comparing the S-dual NCOS configuration (\ref{eq:foten}) with the OD1
configuration given in (\ref{eq:foeight}), we find that all the fields indeed
match exactly with the same identification of parameters 
(\ref{eq:thseventeen}) and
(\ref{eq:theighteen}) along with an extra condition 
\begin{equation}
l= -\tilde l
\label{eq:foeleven}
\end{equation}
This therefore shows that under S-duality the gravity dual configuration of
NCOS theory gets mapped to that of OD1 theory. So, indeed these two theories
are related by strong-weak duality symmetry.

\subsection{On self-duality of (5+1) dimensional NCOS or OD1}

We have already shown in the previous subsection that for the general
case NCOS theory is S-dual to OD1 theory with the parametric relations
$G_o^2 = 1/G_{o(1)}^2$ and $\a'_{\rm eff} = G_{o(1)}^2 \tilde{\a}'_{\rm eff}$.
So the strongly coupled NCOS theory gets mapped to the weakly coupled region
of OD1 theory and vice-versa. However, in order to show the self-duality of
either the (5+1) NCOS theory or OD1 theory we must show that these two
theories may be identified with the parametric relations $G_o = G_{o(1)}$
and $\a'_{\rm eff} = \tilde{\a}'_{\rm eff}/G_{o(1)}^2$. In other words,
the strong coupling region of NCOS (OD1) theory must get mapped to
the strong coupling region of OD1 (NCOS) theory or the weak coupling region 
of one theory must map to the weak coupling region of other theory. We will 
try to see whether at the level of supergravity dual we can make this
identification. However, note that the same OD1 limit (\ref{eq:theleven}) and
(\ref{eq:fofive}) will not do the job and we need to modify it by replacing
$G_{o(1)}^2 \to 1/G_{o(1)}^2$. So, the OD1 limit we take is the following.
\be
\epsilon \to 0, \qquad \cos\psi = \epsilon \to 0, \qquad \sin\varphi
= \tilde{l} \e \to 0,\nonumber
\ee
keeping the following quantities fixed,
\be
\tilde{\a}'_{\rm eff} = \frac{\tilde{\a}'}{\e}, \qquad G_{o(1)}^2 =
\frac{\e}{g_s}, \qquad \tilde{u} = \frac{r}{\e \tilde{\a}'_{\rm eff}},
\label{eq:fotwelve}
\ee
where $\tilde{\a}' = G_{o(1)}^2 \a'$. The harmonic functions (\ref{eq:fosix}) 
in 
this limit take the forms,
\be
H = \frac{1}{b^2 \e^2 \tilde{u}^2}, \qquad H' = \frac{g'}{b^2 \tilde{u}^2},
\qquad H'' = \frac{1}{b^2 \e^2 \tilde{u}^2},
\label{eq:fothirteen}
\ee
where $b^2 = G_{o(1)}^2 \tilde{\a}'_{\rm eff}/N$ and $g' = 1 + 
b^2 \tilde{u}^2$. The supergravity configuration is therefore given as,  
\begin{eqnarray}
&& d\bar{s}^2 = \frac{\e g'^{1/2}}{G_{o(1)}^2}\left[-(d \bar{x}^0)^2 + 
(d\bar{x}^1)^2 + g'^{-1}\sum_{i=2}^5
(d\bar{x}^i)^2 
+ \frac{N \tilde{\a}'_{\rm eff}}{\tilde{u}^2}\left(
d\tilde{u}^2 + \tilde{u}^2 d\Omega_3^2\right)\right], \nonumber \\
&& e^{\bar{\phi}} = \frac{1}{G_{o(1)}^2 b \tilde{u}}, \qquad
\bar{\chi} = \tilde{l}G_{o(1)}^2,\nonumber\\ 
&& \bar{B}_{01} =  \frac{\e^3 \tilde{l}}{G_{o(1)}^2} b^2 \tilde{u}^2,
\qquad \bar{F}_{NS} = \frac{2 N \e \tilde{\a}'_{\rm eff}}{G_{o(1)}^2}\e_3,
\nonumber\\
&& \bar{C}_{01} = \e b^2 \tilde{u}^2, \qquad \bar{F}_{RR} = - 2 N \e
\tilde{l} \tilde{\a}'_{\rm eff} \e_3,\nonumber\\
&&\bar{C}_{2345} = \frac{\e^2}{G_{o(1)}^2 g'},
\label{eq:fofourteen}
\eea
where the fixed coordinates are defined as,
\be
\bar{x}^{0,1} = G_{o(1)} x^{0,1}, \qquad {\rm and} \qquad 
\bar{x}^{2,\ldots, 5} = \frac{G_{o(1)}}{\e} x^{2,\ldots,5}.
\label{eq:fofifteen}
\ee
By comparing (\ref{eq:fofourteen}) with the NCOS supergravity 
configuration (\ref{eq:fofour})
we find that the fields may be identified as the following,
\bea
&& d\bar{s}^2 = ds^2, \qquad e^{\bar{\phi}} = e^{\phi}, \qquad \bar{\chi} =
- \chi, \qquad \bar{B}_{01} = - C_{01}, \qquad \bar{F}_{NS} = - F_{RR}
\nonumber\\
&& \bar{C}_{01} = B_{01}, \qquad \bar{F}_{RR} = F_{NS}, \qquad {\rm and}
\qquad \bar{C}_{2345} = C_{2345},
\label{eq:fosixteen}
\eea
if we impose the following conditions on the parameters of the two theories,
\bea
&& lh''^{1/2} = \frac{1}{G_o^2}, \qquad h' = g', \qquad G_o^2 = G_{o(1)}^2,
\qquad \a'_{\rm eff} = \frac{\tilde{\a}'_{\rm eff}}{G_{o(1)}^2},
\nonumber\\
&& M = N, \qquad {\rm and} \qquad l = -\tilde{l}.
\label{eq:foseventeen}
\eea
Note that the second condition in (\ref{eq:foseventeen}) gives a 
relation between the 
energy parameters of NCOS theory ($u$) and OD1 theory ($\tilde{u}$) as,
\be
u = \tilde{u} G_{o(1)}^3, \qquad {\rm or} \qquad \tilde{u} = \frac{
u}{G_o^3},
\label{eq:foeighteen}
\ee
very similar to the relation we have already found in (\ref{eq:theighteen}). 
From 
(\ref{eq:fosixteen}) and (\ref{eq:foseventeen}), it might seem that we 
have been able to identify
NCOS theory woth OD1 theory implying that we have been able to show the
self-duality conjecture at the level of supergravity dual. However, this
is not true. In fact if we look closely to the first condition in 
(\ref{eq:foseventeen}),
we find that it implies\footnote{If we do not separate the condition as given
below in eq.(\ref{eq:fonineteen}), then this implies that the field
identification (\ref{eq:fosixteen}) can be made only at a single value
of the energy given by $u^2 = \frac{M}{\a'_{\rm eff} G_o^2} - 
\frac{M G_o^2 l^2}{\a'_{\rm eff}}$. Even in this case the effective coupling
$e^\phi$ does not remain small since $e^\phi \ll 1\,\, \Rightarrow\,\, 
u^2 \gg \frac{M}{\a'_{\rm eff} G_o^2}$. So supergravity description can not
be trusted.}
\be
h'' \approx 1, \qquad {\rm and} \qquad l = \frac{1}{G_o^2}.
\label{eq:fonineteen}
\ee
$h'' \approx 1$  implies $\tilde{a}^2 u^2 \ll 1$ or in other words, the
above identification can be made only if the energy parameter in the NCOS
theory satisfies
\be
u^2 \ll \frac{M}{\a'_{\rm eff} G_o^2}.
\label{eq:fotwenty}
\ee
Note that this does not necessarily imply that $h' \approx 1$ also since
$a^2 = \a'_{\rm eff}/(M G_o^2)$. But from the expression of $e^{\phi}$ in
(\ref{eq:fofour}) i.e. for NCOS theory, we find that precisely in the 
energy region
(\ref{eq:fotwenty}), it blows up indicating that the supergravity description
breaks down\footnote{The condition $h'' \approx 1$, $h' \not\approx 1$ can
be satisfied if $G_o^2 \ll 1$ i.e. in the weak coupling region of NCOS
(or OD1) theory. But note that for $G_o^2 \gg 1$, $h''$ and $h'$ get
interchanged and we have $h' \approx 1$, $h'' \not\approx 1$. Thus we still
get the mapping of NCOS and OD1 fields with $h''$ and $h'$ interchanged and
the same conclusion holds for this case.}. This clearly shows that 
the identification of NCOS and OD1
theory can not be made at the level of supergravity dual and therefore in order
to test the self-duality conjecture some non-perturbative techniques may
be required. However, we would like to remark that since we have been able
to map the fields of NCOS theory with those of OD1 theory as expected from
various low energy arguments mentioned in the introduction, it might be
possible that the supergravity description remains valid even in the strong
coupling region (where $e^\phi \gg 1$) due to some underlying 
non-renormalization effect. But we can not justify the validity of such 
remark any further at this point.

\sect{(F, D1, NS5, D5) solution with irrational $\chi_0$, NCOS, OD1
limits and $SL(2,Z)$ duality}

In this section we will study the (F, D1, NS5, D5) supergravity configuration
for irrational $\chi_0$ and obtain both NCOS and OD1 
limits. Since type IIB string theory is believed to possess an $SL(2,Z)$ 
invariance and (F, D1, NS5, D5) state is $SL(2,Z)$ invariant, we also discuss
the $SL(2,Z)$ transformation of both these theories.

Under an $SL(2,Z)$ transformation by the matrix $\left(\begin{array}{cc}
v & w\\ r & s\end{array}\right)$, where $v,\,w,\,r,\,s$ are 
integers with
$vs-rw = 1$, the various fields of type IIB supergravity transform as,
\bea
g_{\mu\nu}^E &\to & g_{\mu\nu}^E, \qquad \tau \to \frac{v\tau + w}
{r\tau + s}, \qquad \left(\begin{array}{c} F_{NS}\\ 
F_{RR}\end{array}\right) \to \left(\begin{array}{cc} s & -r\\ -w & v 
\end{array}\right)\left(\begin{array}{c} F_{NS}\\ F_{RR}\end{array}\right)
\nonumber\\
C_{4} &\to & C_{4},
\label{eq:fione}
\eea
where $g_{\mu\nu}^E$ is the Einstein metric and $\tau = \chi + 
i e^{-\phi}$. The explicit transformation of the dilaton and the axion
are 
\bea
e^{\hat{\phi}} &=& |r\tau + s|^2 e^{\phi} = |\lambda|^2 e^{\phi}\nonumber\\
\hat{\chi} &=& \frac{(v\chi + w)(r\chi + s) + vr 
e^{-2\phi}}{|r\tau + s|^2}
\label{eq:fitwo}
\eea
and the string frame metric $g_{\mu\nu} = e^{\phi/2} g_{\mu\nu}^E$ transforms
as,
\be
d\hat{s}^2 = |\lambda| ds^2,
\label{eq:fithree}
\ee
where we have defined $|\lambda| = |r\tau + s|$. If we demand that the 
transformed metric be asymptotically Minkowskian then
\be
d\hat{s}^2 = \frac{|\lambda|}{|\lambda_0|}
ds^2.
\label{eq:fifour}
\ee
In the above $|\lambda_0| = |r\tau_0 + s| = \left[(r\chi_0+s)^2 + r^2
e^{-2\phi_0}\right]^{1/2}$, with $e^{\phi_0} = g_s$, the closed string coupling
and $\chi_0$, the asymptotic value of the axion. We will use these 
transformations to study the $SL(2,Z)$ duality of the decoupled theories.

\subsection{NCOS limit and $SL(2,Z)$ duality}

The supergravity solution is given in (\ref{eq:twone}), where the harmonic 
functions are now of the forms,
\bea
&& H = 1 + \frac{(M - \chi_0 N) g_s \a'}{\cos\psi \sin\varphi r^2},\nonumber\\
&& H' = 1 + \frac{(M - \chi_0 N) g_s \a' \cos\psi}{\sin\varphi r^2},\nonumber\\
&& H'' = 1 + \frac{(M - \chi_0 N) g_s \a'\cos^2\varphi}
{\cos\psi \sin\varphi r^2}.
\label{eq:fifive}
\eea
In the NCOS decoupling limit (\ref{eq:thfour}) and (\ref{eq:foone}), 
they reduce to
\be
H = \frac{1}{a^2\e^2u^2}, \qquad H' = \frac{h'}{a^2 u^2}, \qquad H'' = 
\frac{h''}{\tilde{a}^2 u^2},
\label{eq:fisix}
\ee
where $a^2 = \a'_{\rm eff}/((M-\chi_0 N) G_o^2)$, $h' = 1 + a^2 u^2$ and 
$\tilde{a}^2 = \a'_{\rm eff}/((M-\chi_0 N) G_o^2 l^2)$, $h'' = 1 + 
\tilde{a}^2 u^2$. Note that the parameters $a$ and $\tilde{a}$ here have
different values from the previous sections. Using (\ref{eq:fione}) and the
decoupling limit (\ref{eq:thfour}) and (\ref{eq:foone}) we get the metric, 
dilaton and 
axion from eq.(\ref{eq:twone}) in the following 
form\footnote{Since the forms of 
the metric and dilaton essentially determine the nature of the decoupled 
theory, we will not give the explicit forms of other fields. The axion is
required for the $SL(2,Z)$ transformation.}
\bea
&& ds^2 = \e l h'^{1/2} h''^{1/2}\left[-(d\tilde{x}^0)^2 + (d\tilde{x}^1)^2
+ h'^{-1}\sum_{i=2}^5(d\tilde{x}^i)^2 + \frac{(M-\chi_0N)\a'_{\rm eff}}{u^2}
(du^2 + u^2 d\Omega_3^2)\right]\nonumber\\
&& e^\phi = \frac{G_o^2 l^2 h''}{au}, \qquad \chi = \frac{1}{h'' G_o^2 l}
+ \chi_0.
\label{eq:fiseven}
\eea
where $\tilde{x}^{0,1,\ldots,5}$ are the same as given in (\ref{eq:thseven}). 
This
is the gravity dual of NCOS theory when $\chi_0$ is irrational. 

In order to
make an $SL(2,Z)$ transformation on this configuration we need to calculate
$|\lambda|$ first. From the forms of $e^\phi$ and $\chi$ in 
(\ref{eq:fiseven}), we
get
\be
|\lambda| = \left(r\chi_0 + s + \frac{r}{G_o^2 l}\right) 
\frac{\hat{h}''^{1/2}}{h''^{1/2}},
\label{eq:fieight}
\ee
where $\hat{h}'' = 1 + \hat{a}^2 u^2$. The parameter $\hat{a}^2$ is defined in
terms of $\tilde{a}^2$ as follows,
\be
\hat{a}^2 = \frac{(r\chi_0 + s)^2}{\left(r\chi_0 + s + 
\frac{r}{G_o^2 l}\right)^2} 
\tilde{a}^2.
\label{eq:finine}
\ee
Substituting $|\lambda|$ (given in (\ref{eq:fieight})) and noting that 
$|\lambda_0|
= (r\chi_0 + s)$ for NCOS limit we find the $SL(2,Z)$ transformed metric from
(\ref{eq:fifour}) as,
\be
d\hat{s}^2 = \e \hat{l} h'^{1/2} \hat{h}''^{1/2}\left[-(d\tilde{x}^0)^2 + 
(d\tilde{x}^1)^2
+ h'^{-1}\sum_{i=2}^5(d\tilde{x}^i)^2 + \frac{(\hat{M}-\hat{\chi}_0\hat{N})
\hat{\a}'_{\rm eff}}{u^2}
(du^2 + u^2 d\Omega_3^2)\right].
\label{eq:fiten}
\ee
Similarly the dilaton can be obtained from (\ref{eq:fitwo}) as,
\be
e^{\hat{\phi}} = \frac{\hat{G}_o^2 \hat{l}^2 \hat{h}''}{au},
\label{eq:fieleven}
\ee
where we have defined the $SL(2,Z)$ transformed parameters $\hat{G}_o^2$,
$\hat{l}$, $\hat{\a}'_{\rm eff}$ and $(\hat{M} - \hat{\chi}_0 \hat{N})$ as
follows,
\bea
&& \hat{G}_o^2 = (r\chi_0 +s)^2 G_o^2, \qquad \hat{l} = \frac{\left(r\chi_0
+ s + \frac{r}{G_o^2 l}\right)}{(r\chi_0 + s)} l\nonumber\\
&& \hat{\a}'_{\rm eff} = (r\chi_0 + s) \a'_{\rm eff}, \qquad
(\hat{M} - \hat{\chi}_0 \hat{N}) = \frac{(M - \chi_0 N)}{(r\chi_0 + s)}.
\label{eq:fitwelve}
\eea
Comparing (\ref{eq:fiten}) and (\ref{eq:fieleven}) with the metric 
and dilaton in 
(\ref{eq:fiseven}) we find that they have precisely the same form and thus we 
conclude that when $\chi_0$ is irrational an $SL(2,Z)$ transformation on
NCOS theory gives another NCOS theory with different parameters related to the
old parameters by eq.(\ref{eq:fitwelve}).

We would like to make a few comments here. First of all, note that the 
scaling parameter $\e$ which is equal to $\cos\psi$ is $SL(2,Z)$ invariant
and so, the coordinates $\tilde{x}^{0,1,\ldots,5}$ have the same form as
the original NCOS gravity dual configuration in (\ref{eq:fiseven}). Also, the
parameter $\tilde{a}^2$ as well as $h''$ transforms under $SL(2,Z)$ 
according to (\ref{eq:finine}) while $a^2$ and $h'$ are $SL(2,Z)$ invariant.
This can be understood from (\ref{eq:twtwo}) since $H$ and $H'$ are $SL(2,Z)$
invariant, but $H''$ is not.
Also note that the parameter $l$ which is proportional to $\cos\varphi$ is
not $SL(2,Z)$ invariant, but transforms according to eq.(\ref{eq:fitwelve}). 
Furthermore, we point out that the combination $(M-\chi_0N)\a'_{\rm eff}$
is $SL(2,Z)$ invariant, but separately they transform according to 
(\ref{eq:fitwelve}).
From the transformation of $\chi_0$, it can be easily checked that if we
start with an irrational $\chi_0$, the transformed NCOS theory will also
have irrational $\chi_0$. The relations between the coupling constants and
the length scales of the two NCOS theories are also given in 
(\ref{eq:fitwelve}).
Lastly, we observe from eq.(\ref{eq:fieight}) that there are two special cases
which may arise. Case (a) $(r\chi_0 + s) = 0$ i.e. $\chi_0$ is rational. In 
this case, the transformed theory (\ref{eq:fiten}) and (\ref{eq:fieleven}) 
reduces to
OD1 theory as expected and studied in section 4. Case (b) $\left(r\chi_0 + s
+ \frac{r}{G_o^2 l}\right) = 0$. In this case, the parameter $l$ of
the transformed theory ($\hat{l}$) vanishes, which means $\cos\varphi$ 
vanishes. This is precisely the NCOS theory we obtained as the decoupling 
limit of (F, D5) configuration studied in subsection 3.1.

\subsection{OD1 limit and $SL(2,Z)$ duality}

We have seen in the previous subsection that when $\chi_0$ is irrational
an NCOS theory will always go over to another NCOS theory with different 
parameters under $SL(2,Z)$ transformation. This will not be true for OD1 
theory as we will see in this subsection. Again we start with the supergravity
configuration in (\ref{eq:twone}). Under the OD1 decoupling limit 
eqs.(\ref{eq:theleven})
and (\ref{eq:fofive}) the harmonic functions take the forms,
\bea
&& H = 1 + \frac{N \a'}{\cos\varphi \cos\psi r^2} = \frac{1}{c^2 \e^2
\tilde{u}^2},\nonumber\\
&& H' = 1 + \frac{\cos\psi N \a'}{\cos\varphi r^2} = \frac{f'}{c^2 
\tilde{u}^2},\nonumber\\
&& H'' = 1 + \frac{\cos\varphi N \a'}{\cos\psi r^2} = \frac{1}{c^2 \e^2 
\tilde{u}^2}.
\label{eq:fithirteen}
\eea
Note that in this case the harmonic functions have exactly the same forms
as in the case of OD1 theory with rational $\chi_0$ given in 
eq.(\ref{eq:foseven}).
Of course, the explicit form of the angle $\cos\varphi$ 
in eq.(\ref{eq:twthree}) is
different in this case. The parameters $c^2$ and $f'$ are as defined before.
The metric, dilaton and the axion in the decoupling limit have the forms,
\bea
&& d\bar{s}^2 = \e  f'^{1/2} G_{o(1)}^2\left[-(d\bar{x}^0)^2 + (d\bar{x}^1)^2
+ f'^{-1}\sum_{i=2}^5(d\bar{x}^i)^2 + \frac{N\tilde{\a}'_{\rm eff}}
{\tilde{u}^2}
(d\tilde{u}^2 + \tilde{u}^2 d\Omega_3^2)\right]\nonumber\\
&& e^{\bar{\phi}} = \frac{G_{o(1)}^2}{c\tilde{u}}, \qquad \bar{\chi} 
= \frac{\tilde{l}}{G_{o(1)}^2}
+ \chi_0.
\label{eq:fifourteen}
\eea
The coordinates $\bar{x}^{0,1,\ldots,5}$ are the same as in 
(\ref{eq:thfourteen}). For
$SL(2,Z)$ transformation we calculate $|\lambda|$ as before and it has the 
form 
\be
|\lambda| = \left(r\chi_0 + s + \frac{r\tilde{l}}{G_{o(1)}^2}\right) f''^{1/2},
\label{eq:fififteen}
\ee
where we have defined
\be
f'' = \left[1 + \frac{r^2}{\left(r\chi_0 + s + \frac{r\tilde{l}}{G_{o(1)}^2}
\right)^2} \frac{c^2 \tilde{u}^2}{G_{o(1)}^4}\right].
\label{eq:fisixteen}
\ee
Now if we define new parameters $l$ and $\tilde{c}$ by the following
relations
\bea
&& r^2 l^2 = \left(r\chi_0 + s + \frac{r\tilde{l}}{G_{o(1)}^2}\right)^2 
G_{o(1)}^4, \nonumber\\
&& \tilde{c}^2 = \frac{c^2}{l^2},
\label{eq:fiseventeen}
\eea
then $f''$ in eq.(\ref{eq:fisixteen}) becomes 
$f'' = 1 + \tilde{c}^2 \tilde{u}^2$ and 
the $SL(2,Z)$ transformd metric and dilaton become
\bea
&& d\hat{s}^2 = \e l f'^{1/2} f''^{1/2}\left[-(d\bar{x}^0)^2 + (d\bar{x}^1)^2
+ f'^{-1}\sum_{i=2}^5(d\bar{x}^i)^2 + \frac{N r\tilde{\a}'_{\rm eff}}
{\tilde{u}^2}
(d\tilde{u}^2 + \tilde{u}^2 d\Omega_3^2)\right],\nonumber\\
&& e^{\hat{\phi}} = f''\frac{r^2 l^2}{G_{o(1)}^2 c\tilde{u}},
\label{eq:fieighteen} 
\eea
where we have scaled the coordinates $\bar{x}^{0,1,\ldots,5} \to \sqrt{r}
\bar{x}^{0,1,\ldots,5}$. Now comparing (\ref{eq:fieighteen}) with the 
metric and dilaton
given in eq.(\ref{eq:fofour}) we notice that if the energy parameters 
$\tilde{u}$ 
and $u$ satisfy the same relation given in (\ref{eq:theighteen}) then 
we get precisely
the NCOS supergravity configuration in (\ref{eq:fofour}). The parameters 
of these 
two theories are related as,
\be
G_o^2 = \frac{r^2}{G_{o(1)}^2}, \qquad {\rm and} \qquad G_o^2 \a'_{\rm eff}
= r \tilde{a}'_{\rm eff}.
\label{eq:finineteen}
\ee
One thing to notice here is that, we started out with an OD1 theory with
irrational $\chi_0$, but after $SL(2,Z)$ transformation we get an NCOS theory
with rational\footnote{This can also be understood from the $SL(2,Z)$
transformation of $\chi_0$ and $g_s$ given in (\ref{eq:fitwo}) for OD1 limit.}
$\chi_0$ (since (\ref{eq:fieighteen}) is the NCOS supergravity 
configuration for rational
$\chi_0$). Also, since NCOS theory with rational $\chi_0$ is S-dual to OD1 
theory with also rational $\chi_0$ (discussed in section 4), so OD1 theory
with irrational $\chi_0$ and rational $\chi_0$ are equivalent.

\sect{(F, D1, NS5, D5) solution, OD5 limit and $SL(2,Z)$ duality}

So far we have seen how NCOS and OD1 theory arise from the decoupling limit
of (F, D1, NS5, D5) supergravity configuration and studied various properties
of them. In this section we will discuss OD5 limit i.e. how OD5 theory also
arises from a decoupling limit of the same supergravity configuration. We
will first consider OD5 limit for rational $\chi_0$ ($\chi_0 = 0$) and study
S-duality and then consider the same limit for irrational $\chi_0$ ($\chi_0
\neq 0$) and study the general $SL(2,Z)$ transformation properties as in the
previous sections.

\subsection{OD5 limit for rational $\chi_0$ and S-duality}

Since $\chi_0$ is rational we can set it to zero by an $SL(2,Z)$ 
transformation. So, the supergravity solution we take is (\ref{eq:twone}) with
$\chi_0 = 0$. OD5 limit is taken by defining a dimensionless scaling parameter
$\e$ with $\cos\varphi = \e \to 0$, keeping the following quantities fixed,
\be
\tilde{\a}'_{\rm eff} = \frac{\a'}{\e}, \qquad G_{o(5)}^2 = \e g_s,
\qquad u = \frac{r}{\e \tilde{\a}'_{\rm eff}}, \qquad \cos\psi = l = 
{\rm finite} < 1.
\label{eq:sione}
\ee
In the above $G_{o(5)}^2$ is the coupling constant of OD5 theory and
$\tilde{\a}'_{\rm eff}$ is the length scale. Note that the scaling
parameter $\e$ and the parameter
$l$ above have nothing to do with the parameters defined in the earlier
sections. Also the case $l = 1$ corresponds to $k = 0$ (see 
eq.(\ref{eq:twthree}))
and so (F, D1) strings are absent and this case has already been studied
before in \cite{gmss,mrone,alioz}. In the decoupling limit (\ref{eq:sione}) the 
harmonic functions
take the forms,
\bea
&& H = 1 + \frac{N \a'}{\cos\varphi \cos\psi r^2} = \frac{1}{d^2 \e^2
u^2},\nonumber\\
&& H' = 1 + \frac{\cos\psi N \a'}{\cos\varphi r^2} = \frac{1}{\tilde{d}^2 \e^2 
u^2},\nonumber\\
&& H'' = 1 + \frac{\cos\varphi N \a'}{\cos\psi r^2} = \frac{F''}{d^2 
u^2},
\label{eq:sitwo}
\eea
where $d^2 = \tilde{\a}'_{\rm eff} l/N$, $\tilde{d}^2 = \tilde{\a}'_{\rm eff}/
(Nl)$ and $F'' = 1 + d^2 u^2$, with $N$ representing the number of NS5 branes.
Therefore, $d\tilde{d} = \tilde{\a}'_{\rm eff}/N$ and $d/\tilde{d} = l$. The
gravity dual configuration of OD5 theory obtained from (\ref{eq:twone}) is
\begin{eqnarray}
&& ds^2 = \e F''^{1/2}\left[-(d \tilde{x}^0)^2 + 
\sum_{i=1}^5 (d\tilde{x}^i)^2 
+ \frac{N \tilde{\a}'_{\rm eff}}{u^2}\left(
du^2 + u^2 d\Omega_3^2\right)\right], \nonumber \\
&& e^\phi = \frac{G_{o(5)}^2 F''}{du}, \qquad
\chi = \frac{1}{G_{o(5)}^2 F''}, \nonumber\\
&& B_{01} = \frac{\sqrt{1-l^2}}{l}\e^2 d^2 u^2,
\qquad F_{NS} = 2 N \a'\e_3, \nonumber\\
&& C_{01} = \frac{\e^4 \sqrt{1-l^2}}{l G_{o(5)}^2} d^2 u^2, \qquad 
F_{RR} = - 2 M \a'\e_3,\nonumber\\ 
&& C_{2345} = \frac{\e l \sqrt{1-l^2}}{G_{o(5)}^2},
\label{eq:sithree}
\eea
where the fixed coordinates are defined as
\be
\tilde{x}^{0,1} = \sqrt{l} x^{0,1}, \qquad \tilde{x}^{2,\ldots,5} = \frac{1}
{\sqrt{l}} x^{2,\ldots,5},
\label{sifour}
\ee
where $M$ is the number of D5 branes. This is the gravity dual description of
OD5 theory. The S-dual configuration of OD5 theory can be obtained using the 
general relations given in (\ref{eq:fonine}). We calculate 
$|\lambda| = [\chi^2
+ e^{-2\phi}]^{1/2}$ from (\ref{eq:sithree}) for this purpose as,
\be
|\lambda| = \frac{1}{F''^{1/2} G_{o(5)}^2}.
\label{eq:sifive}
\ee
Also insisting that the transformed metric be asymptotically Minkowskian such
that $d\hat{s}^2 = (|\lambda|/|\lambda_0|) ds^2$, with $|\lambda_0| = 
\e/G_{o(5)}^2$, we get the S-dual configuration as,
\begin{eqnarray}
&& d\hat{s}^2 = -(d \tilde{x}^0)^2 + 
\sum_{i=1}^5 (d\tilde{x}^i)^2 
+ \frac{N \tilde{\a}'_{\rm eff}}{u^2}\left(
du^2 + u^2 d\Omega_3^2\right), \nonumber \\
&& e^{\hat{\phi}} = \frac{1}{G_{o(5)}^2 du}, \qquad
\hat{\chi} = - G_{o(5)}^2 = - \frac{N}{M} = {\rm a\,\, rational\,\, no.}, 
\nonumber\\
&& \hat{B}_{01} = -C_{01}, \qquad 
\hat{F}_{NS} = - F_{RR},
\nonumber\\
&& \hat{C}_{01} = B_{01}, \qquad 
\hat{F}_{RR} =  F_{NS},\nonumber\\ 
&& \hat{C}_{2345} = C_{2345}.
\label{eq:sisix}
\eea
We note from above that in the UV, $e^{\hat{\phi}} \ll 1$ and for $N \gg 1$,
the curvature remains small and therefore we have a valid supergravity
description. By comparison \cite{abks,imsy}, we find that this is precisely the
supergravity dual of LST, where the closed string coupling $g_s = 
\e/G_{o(5)}^2 \to 0$ and $\tilde{\a}'_{\rm ef} =$ finite (the length scale of
LST). Thus we conclude that OD5 theory goes over to LST under type IIB
S-duality.

\subsection{OD5 limit for irrational $\chi_0$ and $SL(2,Z)$ duality}

In this case $\chi_0 \neq 0$ in the (F, D1, NS5, D5) bound state given in
(\ref{eq:twone}). Under OD5 limit (\ref{eq:sione}), the harmonic 
functions take exactly 
the same form as given in (\ref{eq:sitwo}), although the explicit form of 
$\cos\varphi$ is different for this case. So, the metric and dilaton have 
exactly the same form as in (\ref{eq:sithree}), but the axion is 
modified to include
$\chi_0$ term\footnote{Here also we do not give the explicit forms of other
gauge fields since we will only indicate the nature of the $SL(2,Z)$ 
transformed theory.}. So,
\be
\chi = \frac{1}{G_{o(5)}^2 F''} + \chi_0.
\label{eq:siseven}
\ee
The $SL(2,Z)$ transformation of various fields are given in 
(\ref{eq:fione}) --
(\ref{eq:fifour}). For this purpose we calculate,
\be
|\lambda| = |r\tau + s| = \left(r\chi_0 + s + \frac{r}{G_{o(5)}^2}\right)
\frac{\hat{F}''^{1/2}}{F''^{1/2}},
\label{eq:sieight}
\ee
where $\hat{F}'' = 1 + \hat{d}^2 u^2$. The parameter $\hat{d}^2$ is given in 
terms of $d^2$ as,
\be
\hat{d}^2 = \frac{(r\chi_0 + s)^2}{\left(r\chi_0 + s + \frac{r}
{G_{o(5)}^2}\right)^2} d^2.
\label{eq:sinine} 
\ee
The $SL(2,Z)$ transformed metric and dilaton can be obtained from 
(\ref{eq:fifour})
and (\ref{eq:fitwo}) as, 
\begin{eqnarray}
&& d\hat{s}^2 = \hat{\e} \hat{F}''^{1/2}\left[-(d \tilde{x}^0)^2 + 
\sum_{i=1}^5 (d\tilde{x}^i)^2 
+ \frac{\hat{N} \hat{\tilde{\a}}'_{\rm eff}}{u^2}\left(
du^2 + u^2 d\Omega_3^2\right)\right], \nonumber \\
&& e^{\hat{\phi}} = \frac{\hat{G}_{o(5)}^2 \hat{F}''}{\hat{d}u},
\label{eq:siten}
\eea
where the $SL(2,Z)$ transformed parameters are given as,
\bea
&& \hat{\e} = \frac{\left(r\chi_0
+ s + \frac{r}{G_{o(5)}^2}\right)}{(r\chi_0 + s)} \e, \qquad
\hat{\tilde{\a}}'_{\rm eff} = \frac{(r\chi_0 + s)}
{\left(r\chi_0
+ s + \frac{r}{G_{o(5)}^2}\right)} \tilde{\a}'_{\rm eff},\nonumber\\ 
&& \hat{N} = \frac{\left(r\chi_0
+ s + \frac{r}{G_{o(5)}^2}\right)}{(r\chi_0 + s)} N, \qquad
\hat{G}_{o(5)}^2 = 
\left(r\chi_0
+ s + \frac{r}{G_{o(5)}^2}\right)(r\chi_0 + s) G_{o(5)}^2.
\label{eq:sieleven} 
\eea
Comparing (\ref{eq:siten}) with the metric and dilaton in (\ref{eq:sithree}), 
we 
find that they have exactly the same form and therefore we conclude that
for $\chi_0$ = irrational an OD5 theory goes over to another OD5 theory
under $SL(2,Z)$ transformation. Notice that since $l = \cos\psi$ is
$SL(2,Z)$ invariant the coordinates $\tilde{x}^{0,1,\ldots,5}$ remain the 
same in the two OD5 theories. However, the scaling parameter 
$\e = \cos\varphi$
transforms according to (\ref{eq:sieleven}). Also, 
since $H$ and $H'$ are $SL(2,Z)$
invariant, so both $d$ and $\tilde{d}$ must transform in the same way as
given in (\ref{eq:sinine}) and $\e$ must transform in the 
opposite way to $d$ and
$\tilde{d}$ as obtained in (\ref{eq:sieleven}). The combination 
$N\tilde{\a}'_{\rm eff}$ is $SL(2,Z)$ invariant. The coupling constant 
$G_{o(5)}^2$ and the length scale $\tilde{\a}'_{\rm eff}$ of the $SL(2,Z)$
transformed OD5 theory are given in (\ref{eq:sieleven}). From 
eq.(\ref{eq:sieight}) we 
find that when $r\chi_0 + s = 0$ i.e. when $\chi_0$ is rational, the $SL(2,Z)$
transformed configuration (\ref{eq:siten}) reduces precisely to LST studied in
the previous subsection as expected. However, $r\chi_0 + s + r/G_{o(5)}^2$ 
can not be zero in this case because that would imply $\hat{\e}$ to be exactly
equal to zero and there would be no scaling parameter to obtain a 
decoupled theory.

\sect{Conclusion}

To summarize, starting from an $SL(2,Z)$ invariant bound state (F, D1, NS5, D5)
of type IIB supergravity we have shown how to take various decoupling limits
leading to the supergravity duals of NCOS theory, OD1 theory and OD5 theory.
The decoupled theories obtained in this way are in general different from those
obtained  by taking low energy limits on (F, D5), (D1, NS5) and (NS5, D5)
separately in the sense that their field contents are different. As a result,
the decoupled theories or their supergravity duals can be subjected to the
full $SL(2,Z)$ duality of type IIB supergravity. We have studied the $SL(2,Z)$
transformation properties of various (5+1) dimensional decoupled theories
obtained from this bound state. The asymptotic value of the RR scalar or
axion present in the supergravity configuration is crucial to determine
the behavior of the decoupled theories under the $SL(2,Z)$ transformation.
We found that when $\chi_0$ is rational the general NCOS and OD1 theories
are related by the strong-weak duality as obtained before in the special
cases i.e. for (F, D5) and (D1, NS5) solutions in \cite{gmss,hartwo}. But
when $\chi_0$ is irrational NCOS theory gives another NCOS theory with
different values of the parameters. However, for OD1 theory we 
surprisingly found a different conclusion that even for irrational $\chi_0$,
OD1 theory goes over to an NCOS theory under $SL(2,Z)$ transformation. This
shows that OD1 theories with rational and irrational $\chi_0$ are equivalent.
We have also addressed the question of self-duality of both (5+1) dimensional
NCOS and OD1 theories in this context. We have been able to show the 
self-duality only in the region where the coupling constant blows up and the
supergravity description breaks down. We mentioned that this result can be
interpreted in two ways. It can either mean that the self-duality conjecture
can not be tested at the level of supergravity dual or it can be taken as
a supporting evidence for the self-duality conjecture if the supergravity
description somehow remains valid in the strongly coupled region due to some
underlying non-renormalization effect. For completeness, we have also studied
the $SL(2,Z)$ transformation on OD5 theories. Here we found that for rational
$\chi_0$, OD5 is related to LST by the S-duality of type IIB theory, but
for irrational $\chi_0$, OD5 theory gives another OD5 theory with different
values of the parameters. In conclusion, we point out that since type IIB
string theory is believed to possess an $SL(2,Z)$ symmetry, by working with
the $SL(2,Z)$ invariant supergravity configuration, we have been able to access
the full $SL(2,Z)$ group of transformation to relate various decoupled 
theories in (5+1) dimensions.

\section*{Acknowledgements}

We would like to thank Jianxin Lu for numerous discussions and for
participation at an early stage of this work.

%\newpage

\end{document}